\documentclass[aps,prd,superscriptaddress]{revtex4}
\usepackage{epsfig,epsf}
\usepackage{amsmath}
\usepackage{amsthm}
\usepackage{amsfonts}
\usepackage{amssymb}
\usepackage{dsfont}
\usepackage{multirow}
\usepackage{appendix}
\usepackage{slashed}
\usepackage[active]{srcltx}
\usepackage{psfrag}

\begin{document}

\title{Nature of the $\Omega(2012)$ through its strong decays} 
\date{\today}
\author{T.~M.~Aliev}
\affiliation{Physics Department,
Middle East Technical University, 06531 Ankara, Turkey}
\author{K.~Azizi}
\affiliation{Physics Department, Do\u gu\c s University,
Ac{\i}badem-Kad{\i}k\"oy, 34722 Istanbul, Turkey}
\affiliation{School of Physics, Institute for Research in Fundamental Sciences (IPM), P. O. Box 19395-5531, Tehran, Iran}
\author{Y.~Sarac}
\affiliation{Electrical and Electronics Engineering Department,
Atilim University, 06836 Ankara, Turkey}
\author{H.~Sundu}
\affiliation{Department of Physics, Kocaeli University, 41380 Izmit, Turkey}

\begin{abstract}
We extend our previous analysis on the mass of the recently discovered $\Omega(2012)$ state by investigation of its strong  decays and  calculation of its  width employing the method of light cone QCD sum rule. Considering two possibilities for the quantum numbers of $\Omega(2012)$ state, namely $1P$ orbital excitation with $J^P=\frac{3}{2}^-$ and $2S$ radial excitation with $J^P=\frac{3}{2}^+$, we obtain the strong coupling constants defining the  $\Omega(1P/2S)\rightarrow\Xi K$ decays. The results of the coupling constants are then used to calculate  the  decay width corresponding to each possibility. Comparison  of the obtained results on the total widths in this work with the experimental value  and taking into account the results of our previous mass prediction on the $\Omega(2012)$ state, we conclude that this state is  $1P$ orbital excitation of the ground state $\Omega$ baryon, whose quantum numbers are $J^P=\frac{3}{2}^-$.

\end{abstract}

\maketitle
\section{Introduction}
The study of the different parameters of the hadrons  allows us  gain deeper understanding on the properties of the observed particles as well as provide insight into the future experiments searching for the new states. Up to now many baryons have been observed and their properties have been examined extensively. However we are still in need of much work to obtain comprehensive information even for the light baryons. We need more information especially on their excited states. For these baryons some excited states predicted by the quark model have not been observed yet experimentally. Therefore investigations on these experimentally not yet observed states, providing information to the experimental researches, are essential.   

Among these states are the excited states of the $\Omega$ with three strange quark content. So far the list provided by the Particle Data Group (PDG)~\cite{Tanabashi2018} contains a few of the $\Omega$ states. Except for the ground state $\Omega(1672)$, our knowledge is limited on the nature of these baryons. Therefore the recent observation of the Belle Collaboration reporting $\Omega(2012)$ state with mass  $2012.4 \pm 0.7$~(stat)~$\pm 0.6$(syst)~MeV and width $6.4^{+2.5}_{-2.0}$~(stat)~$\pm 1.6$~(syst)~MeV~\cite{Yelton:2018mag} has triggered the attentions of the researchers on this particle. The Belle Collaboration observed this particle in the $\Omega^{*-}\rightarrow \Xi^0K^-$ and $\Omega^{*-}\rightarrow \Xi^{-}K_S^0$ decays and reported its being an excited $\Omega^-$ state with probable quantum numbers $J^P=\frac{3}{2}^-$. This decision was made based on the comparison of the observation with the present theoretical works on the masses of excited $\Omega$ states with different models. The mass predictions of these models for $J^P=\frac{3}{2}^-$ state are close to the observed value. Among these models are the quark model~\cite{Chao:1980em,Kalman:1982ut,Capstick:1986bm,Loring:2001ky,Liu:2007yi,Pervin:2007wa,An:2013zoa,An:2014lga,Faustov:2015eba}, lattice gauge theory~\cite{Engel:2013ig,Liang:2015bxr} and Skyrme model~\cite{Oh:2007cr}.  The predictions of these models put support behind the possibility of $J^P=\frac{3}{2}^-$ assignment for the observed state. However besides the mass prediction, the investigation of other properties of the considered state would be helpful to identify the state more reliably. The magnetic dipole moment and Radiative decays are among these properties which may help us gain information about the properties of these particles. Such investigations for negative parity baryons exist in the literature~\cite{Aliev:2014pfa,Aliev:2015,Kaxiras:1985zv}. The study of the strong decay of the particle is also helpful in this respect. With this motivation, to provide a possible interpretation for the observed $\Omega(2012)$ particle, using the chiral quark model an analysis on its strong decay was carried out in Ref.~\cite{Xiao:2018pwe}. The result of this work suggested the possibility of $\Omega(2012)$ being $1P$ state with $J^P=\frac{3}{2}^-$ without completely excluding the other possibilities such as $1P$ state with $J^P=\frac{1}{2}^-$ and $2S$ state with $J^P=\frac{3}{2}^+$ quantum numbers. The possibility of the $\Omega(2012)$ being a hadronic molecular state also discussed in Refs.~\cite{Polyakov:2018mow,Valderrama:2018bmv,Lin:2018nqd}.       

Radial excitation of the decuplet baryons were investigated in Refs.~\cite{Chao:1980em} and \cite{Aliev:2016jnp} and mass value for radial excitation of $\Omega$ state was predicted as 2065~MeV and $2176\pm 219$~MeV, respectively. In Ref.~\cite{Aliev:2018syi} the mass of the newly observed $\Omega(2012)$ state was extracted using QCD sum rule approach. Giving very consistent mass value with that of the experimentally observed $\Omega(2012)$ state, this result allowed us to interpret this state as the orbital excitation of the ground state $\Omega$ baryon. To understand the nature of $\Omega(2012)$, the investigations on other aspects, namely the strong decay of the $\Omega(2012)$ state into $\Xi^0$ and $K^-$, is necessary. Considering this, in this work we extend our analysis for $\Omega(2012)$ presented in Ref.~\cite{Aliev:2018syi} using the spectroscopic parameters obtained in~\cite{Aliev:2018syi} as input parameters. We investigate the $\Omega(2012)\rightarrow\Xi^0K^-$ transition and calculate corresponding coupling constant. While doing so, we consider two possibilities for the $\Omega(2012)$ taking it as being $1P$ or $2S$ state with spin-parity quantum numbers $J^P=\frac{3}{2}^-$ and $J^P=\frac{3}{2}^+$, in what follows represented by $\widetilde{\Omega}$ and $\Omega^{\prime}$ respectively, and calculate the decay widths for these possible configurations. Then we compare the results of these analysis with the experimental value of the width. For the calculations we apply the light cone QCD sum rule (LCSR) approach ~\cite{Braun:1988qv, Balitsky:1989ry,Chernyak:1990ag} which is an extension of the traditional QCD sum rule method. This method has been extensively and successfully used to study the various properties of the hadrons such as form factors, strong coupling constants etc. 

The arrangement of the remaining part of the paper is as follows: Section II presents the calculations of the coupling constants for  $\widetilde{\Omega}$ and $\Omega^{\prime}$ transitions to  $\Xi^0K^-$ in detail. Section III is devoted to the numerical analysis of the results and decay width calculations employing the results obtained for coupling constants. Final section is separated for the conclusion. In the last section the experimental width is also compared with the results of the decay widths obtained for different $J^P$ scenarios assigned to the $\Omega(2012)$ state.

\section{ $\widetilde{\Omega}$ and $\Omega^{\prime}$ transitions to  $\Xi^0K^-$ } 

The calculation of the strong coupling constants $g_{\widetilde{\Omega}\Xi K}$ and $g_{\Omega^{\prime}\Xi K}$ corresponding to the decays $\widetilde{\Omega}\rightarrow \Xi^0K^-$  and $\Omega^{\prime}\rightarrow \Xi^0K^-$, respectively, with the light cone QCD sum rules (LCSR) are presented in this section. For the calculation of the coupling constants we start with a correlation function having the following form:   
\begin{equation}
\Pi _{\mu}(q)=i\int d^{4}xe^{iq\cdot x}\langle K(q)|\mathcal{T}\{J_{\Xi
}(x)\bar{J}_{\Omega\mu }(0)\}|0\rangle.  \label{eq:CorrF1}
\end{equation}
The interpolating currents of $\Omega$ and $\Xi$ baryons, $J_{\Omega\mu}$ and $J_{\Xi}$, are written in terms of quark fields as: 
\begin{eqnarray}\label{Eq:Current2}
J_{\Omega\mu}&=& \epsilon^{abc} (s^{aT}C\gamma_\mu s^{b})s^{c},
\end{eqnarray}
and
\begin{eqnarray}\label{Eq:Current3}
J_{\Xi} &=&\epsilon^{abc}
\Big\{\Big(s^{T,a}(x)Cu^b(x)\Big)\gamma_5 s^c(x)
+\beta\Big(s^{T,a}(x)C\gamma_5u^b(x)\Big)s^c(x) \Big\}~,
\end{eqnarray}
where $a$, $b$ and $c$ are used to represent the color indices, $\beta$ is an arbitrary parameter and $C$ is the charge conjugation operator.

The correlation function can be calculated either in terms of hadronic degrees of freedom or in terms of QCD degrees of freedom. In QCD sum rule formalism  firstly we get these two representations of the correlator. Its representation acquired in terms of hadronic parameters such as mass, residue, coupling constant of considered hadrons  is called as the physical or phenomenological side. The second representation including QCD parameters such as masses of quarks, quark-gluon condensates is called QCD or theoretical side. By matching of the coefficients of the same Lorentz structures obtained in both sides we obtain the QCD sum rules for the physical parameters in question.   

We firstly focus on the case $J^P=\frac{3}{2}^-$, orbital excitation of the ground state $\Omega$, and investigate the $\widetilde{\Omega}\rightarrow \Xi^0K^-$ transition. Inserting complete sets of hadronic states carrying the same quantum numbers with the hadrons of interest we get the hadronic representation of the calculations. The result of this side for orbital excitation $\widetilde{\Omega}$ has the following form:
\begin{eqnarray}
\Pi ^{\mathrm{Phys}}_\mu(p,q)=\frac{\langle 0|J _{\Xi}|\Xi (p,s)\rangle
}{p^{2}-m_{\Xi}^{2}}\langle K(q)\Xi(p,s)|\widetilde{\Omega}
(p^{\prime },s^{\prime })\rangle  \frac{\langle \widetilde{\Omega}(p^{\prime },s^{\prime
})|\bar{J}_{\Omega\mu}|0\rangle }{p^{\prime
2}-m_{\widetilde{\Omega}}^{2}}
 +\ldots ,  \label{eq:SRDecay}
\end{eqnarray}
%
%
where the $\ldots $ represents the contribution of higher states and continuum. In the last equation the $p^{\prime}=p+q$, $p$ and $q$ are the momenta of $\widetilde{\Omega}$, $\Xi$ and $K$ states, respectively. The result comes up with the matrix elements which can be written in terms of physical parameters as
\begin{eqnarray}
\langle 0|J_{\Omega\mu}|\widetilde{\Omega}(p^{\prime},s^{\prime})\rangle
&=&\lambda_{\widetilde{\Omega}}\gamma_5 u_\mu(p^{\prime},s^{\prime}),
\nonumber \\
\langle 0|J _{\Xi }|\Xi(p,s)\rangle & =&\lambda_{\Xi}
u(p,s), \nonumber \\
\langle K(q)\Xi (p,s)|\widetilde{\Omega} (p^{\prime },s^{\prime
})\rangle &=& g_{\widetilde{\Omega} \Xi
K}\overline{u}(p,s)\gamma_5 u_{\mu}(p^{\prime },s^{\prime })q^{\mu},
  \label{eq14}
\end{eqnarray}
where $\lambda_i$ with $i=\widetilde{\Omega},~\Xi$ is the residue of the related baryon, $g_{\widetilde{\Omega} \Xi K}$ represents the coupling constant and $u_{\mu}$ is the Rarita-Schwinger bispinor for spin-$\frac{3}{2}$ state.

Together with these matrix elements and performing the summation over spins of the $\Xi$ and $\widetilde{\Omega}$ baryons by using
\begin{eqnarray}
\sum_{s}u(p,s)\bar{u}(p,s)&=&({\slashed
p}+m_{\Xi}),\nonumber\\
\sum_{s'}u_{\mu}(p',s')\bar{u}_{\nu}(p',s')&=&-({\slashed
p'}+m_{\widetilde{\Omega}})\left[ g_{\mu\nu}-\frac{1}{3}\gamma_{\mu} \gamma_{\nu}-
\frac{2p'_{\mu}p'_{\nu}}{3m_{\widetilde{\Omega}}^2}
+\frac{p'_{\mu}\gamma_{\nu}-p'_{\nu}\gamma_{\mu}}{3m_{\widetilde{\Omega}})}\right],
\label{eq:SumPc}
\end{eqnarray}
for the hadronic side of the correlation function we get
\begin{eqnarray}
\Pi _{\mu }^{\mathrm{Phys}}(p,q)=\frac{g_{\widetilde{\Omega} \Xi K}\lambda_{\Xi}\lambda_{\widetilde{\Omega}}}{(p^{2}-m_{\Xi }^{2})(p^{\prime }{}^{2}-m_{\widetilde{\Omega}}^{2})}q^{\nu }(\slashed p+m_{\Xi})\gamma _{5}  \left( \slashed p^{\prime}+m_{\widetilde{\Omega}}\right) T_{\nu \mu }(%
m_{\widetilde{\Omega}})\gamma _{5}+\ldots,
\end{eqnarray}
%
%
%
where $T_{\nu\mu}$ is used to represent the expression
\begin{eqnarray}
&&T_{\nu \mu }(m)=g_{\nu \mu }-\frac{1}{3}\gamma _{\nu }\gamma
_{\mu }-\frac{2}{3m^{2}}p'_{\nu }p'_{\mu }+\frac{1}{3m}\left( p'_{\nu }\gamma _{\mu }-p'_{\mu }\gamma _{\nu }\right) .
\end{eqnarray}%
After the double Borel transformation with respect to the variables $-p^2$ and $-p^{\prime}{}^2$  we get the final result for the physical side as
\begin{eqnarray}\label{PiPHYS}
\mathcal{B}\Pi _{\mu }^{\mathrm{Phys}}(p,q)=g_{\widetilde{\Omega}\Xi K}\lambda _{\Xi} \lambda_{\widetilde{\Omega}} e^{-m_{\widetilde{\Omega}}%
^{2}/M_{1}^{2}}e^{-m_{\Xi}^{2}/M_{2}^{2}}q^{\nu }(\slashed p+m_{\Xi})\gamma _{5}\left( \slashed p^{\prime}+%
m_{\widetilde{\Omega}}\right) T_{\nu \mu }(m_{\widetilde{\Omega}})\gamma _{5}, 
\end{eqnarray}
where $q^2=m_K^2$ and $m_K$ is the mass of the $K$-meson. $M_1^2$ and $M_2^2$ represents the Borel parameters. As is seen from the above equation there are many structures entering the calculations. Among these structures we need only the ones that are independent and give contributions only to the spin-3/2 states at initial channel. The current $J_{\Omega\mu} $  in Eq. (\ref{Eq:Current2}) couples not only to the spin-3/2 states but also to the possible spin-1/2 states with the same quark contents. Hence we need to remove the spin-1/2 pollution to get pure spin-3/2 contributions. The procedure for removing these unwanted contributions is presented for instance in Ref. \cite{Aliev:2010su} in details. After applying this procedure there remain only four structures, $\slashed q q_{\mu}$, $\slashed p q_{\mu}$, $\slashed q \slashed p q_{\mu}$ and $ q_{\mu}$ that give contributions to the pure spin-3/2 states at initial channel. We will use these structures and match their coefficients in physical side to the ones obtained in the QCD side of the calculations.
%
%
%

The QCD side of the correlation function is made with the same correlation function given in Eq.~(\ref{eq:CorrF1}) using the explicit form of interpolating currents. After inserting the currents to the correlation function we make the contraction via Wick's theorem considering all possible contractions between the quark fields. The contractions give us the result in terms of the light quark propagators in the presence of background field which contain the perturbative and non-perturbative terms. The propagators are used explicitly in the coordinate space and via Fourier transformation the expression is converted to the momentum space. In addition to the propagators, in the calculations there appear matrix elements of non-local operators between $K$-meson and vacuum states which have the common form $\langle K(q)|\bar{q}(x)\Gamma q(y)|0\rangle$ or $\langle K(q)|\bar{q}(x)\Gamma G_{\mu\nu} q(y)|0\rangle$. These matrix elements are parameterized in terms of $K$-meson distribution amplitudes (DAs). The $\Gamma$ in these matrix elements is the full set of Dirac matrices and $G_{\mu\nu}$ is the gluon field strength tensor.  These matrix elements are used as inputs of LCSR to get the nonperturbative contributions. The DAs for $K$-meson are derived in Refs.~\cite{Ball:2006wn,Belyaev:1994zk,Ball:2004ye} and their expressions can be found there. 

To suppress the contribution of higher states and continuum the Borel transformation is also applied to this side of the correlation function. Final result is obtained after continuum subtraction using quark-hadron duality assumption. Since this side of calculation ends up with very lengthy expressions and we don't want to load the text with lengthy equations, we shall skip writing these results. Here we chose the coefficients of the same Lorentz structures as in physical side which can be represented as
%
%
\begin{eqnarray}
\mathcal{B}\Pi _{\mu }^{\mathrm{QCD}}(p,q)\equiv\mathcal{B}\Pi_1^{QCD}\slashed q q_{\mu}+ \mathcal{B}\Pi_2^{QCD}\slashed p q_{\mu}+\mathcal{B}\Pi_3^{QCD}\slashed q \slashed p q_{\mu}+\mathcal{B}\Pi_4^{QCD} q_{\mu}. 
\end{eqnarray}
%
Separating the coefficients of the selected Lorentz structures in both sides and matching these results we get the QCD sum rules for the coupling constants under consideration. For the structure $ \slashed q q_{\mu} $, for instance,  and the $\widetilde{\Omega}\rightarrow \Xi^0K^-$ transition we get  
\begin{eqnarray}
g_{\widetilde{\Omega}\Xi K}=\frac{e^{m_{\widetilde{\Omega}}%
^{2}/M_{1}^{2}}e^{m_{\Xi}^{2}/M_{2}^{2}}}{\lambda_{\widetilde{\Omega}}\lambda _{\Xi}m_{\Xi}}\mathcal{B}\Pi_1^{QCD}.
\end{eqnarray}
The result obtained above includes two Borel parameters $M_1^2$ and $M_2^2$. Using the fact that the masses of $\widetilde{\Omega}$ and $\Xi$ are close to each other, we can choose
\begin{eqnarray}
M_1^2=M_2^2=2M^2.
\end{eqnarray}

For the case when the $\Omega(2012)$ is considered as radial excitation of the ground state $\Omega$ the coupling constant for the transition $\Omega^{\prime}\rightarrow \Xi^0K^-$ can be obtained from the result of the coupling constant obtained for $\widetilde{\Omega}\rightarrow \Xi^0K^-$. To this end, it suffices to make the replacements $m_{\widetilde{\Omega}}\rightarrow -m_{\Omega^{\prime}}$ and $\lambda_{\widetilde{\Omega}}\rightarrow \lambda_{\Omega^{\prime}}$. 

\section{Numerical analysis}
For performing the numerical analysis of the obtained sum rules for the coupling constants besides the $K$-meson distribution amplitudes given in Refs.~~\cite{Ball:2006wn,Belyaev:1994zk,Ball:2004ye} we need the values of various input parameters, such as quark condensates, masses and residues of the considered hadrons, which are presented in table~\ref{tab:Param1}.
\begin{table}[tbp]
\begin{tabular}{|c|c|}
\hline\hline
Parameters & Values \\ \hline\hline
$m_{\Xi}$                               & $1314.86\pm 0.20~\mathrm{MeV}$ \cite{Tanabashi2018}\\
$m_{\widetilde{\Omega}}$                & $2019^{+17}_{-29}~\mathrm{MeV}$ \cite{Aliev:2018syi}\\
$m_{\Omega^{\prime}}$                   & $2176\pm219~\mathrm{MeV}$ \cite{Aliev:2016jnp}\\
$\lambda_{\Xi}(1\mbox{GeV})$                         & $0.017\pm 0.003~\mathrm{GeV}^3$ \cite{Azizi:2015ica}\\
$\lambda_{\widetilde{\Omega}}(1\mbox{GeV})$          & $0.108^{+0.004}_{-0.005}~\mathrm{GeV}^3$ \cite{Aliev:2018syi}\\
$\lambda_{\Omega^{\prime}}(1\mbox{GeV})$             & $0.129\pm 0.039~\mathrm{GeV}^3$ \cite{Aliev:2016jnp}\\
$f_{K}(1\mbox{GeV})$                                 & $160~\mathrm{MeV}$ \cite{Ball:2004ye}\\
$m_{s}(1\mbox{GeV})$                                  & $128^{+12}_{-4}~\mathrm{MeV}$ \cite{Tanabashi2018}\\
$\langle \bar{q}q \rangle (1\mbox{GeV})$& $(-0.24\pm 0.01)^3$ $\mathrm{GeV}^3$ \cite{Belyaev:1982sa}  \\
$\langle \bar{s}s \rangle (1\mbox{GeV})$             & $0.8\langle \bar{q}q \rangle$ \cite{Belyaev:1982sa} \\
$m_{0}^2 (1\mbox{GeV})$                              & $(0.8\pm0.1)$ $\mathrm{GeV}^2$ \cite{Belyaev:1982sa}\\
$\langle g_s^2 G^2 \rangle $            & $4\pi^2 (0.012\pm0.004)$ $~\mathrm{GeV}^4 $\cite{Belyaev:1982cd}\\
$ \Lambda (1\mbox{GeV})$                             & $ (0.5\pm0.1) $ $\mathrm{GeV} $ \cite{Chetyrkin:2007vm} \\
\hline\hline
\end{tabular}%
\caption{Some input parameters.}
\label{tab:Param1}
\end{table}

\begin{figure}[h!]
\begin{center}
\includegraphics[totalheight=5cm,width=7cm]{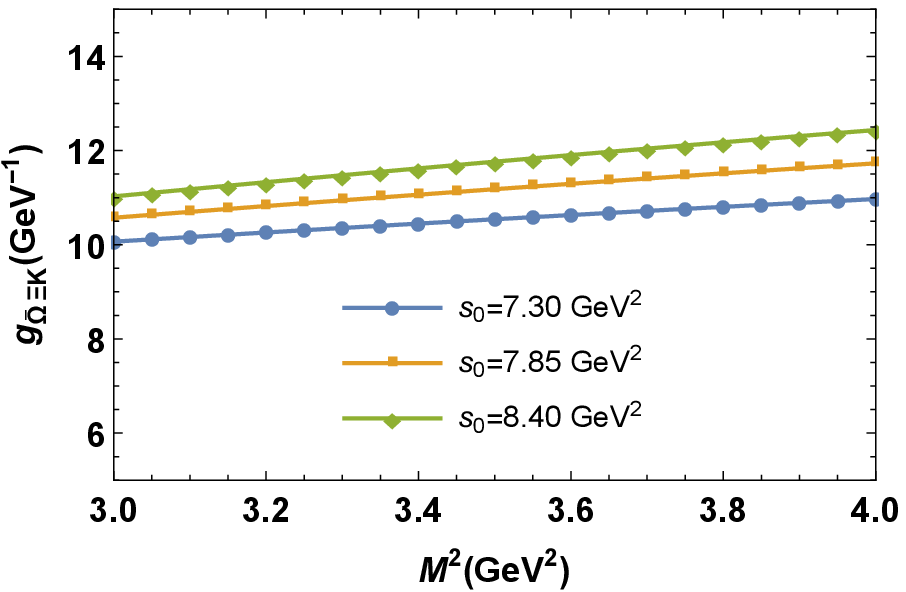}
\includegraphics[totalheight=5cm,width=7cm]{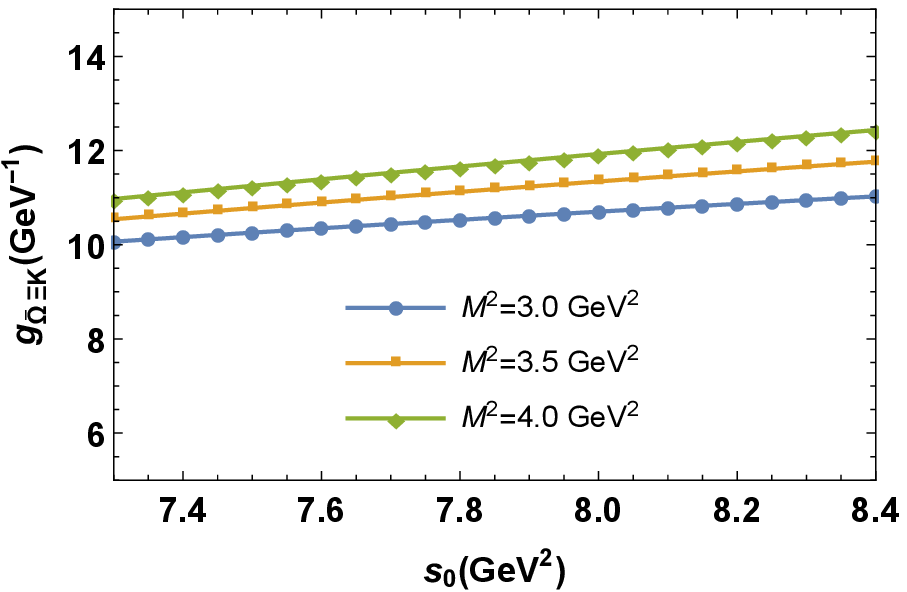}
\end{center}
\caption{\textbf{Left:} The coupling constant $g_{\widetilde{\Omega}\Xi K} $ for the transition of the orbitally excited $\Omega$  baryon vs Borel
parameter $M^2$ for the structure $ \slashed q q_{\mu} $.
\textbf{Right:} The coupling constant $g_{\widetilde{\Omega}\Xi K} $ for the transition of the orbitally excited $\Omega$   baryon vs threshold
parameter $s_0$ for the structure $ \slashed q q_{\mu} $. }
\label{gr:gOmega1PXiK}
\end{figure}

\begin{figure}[h!]
\begin{center}
\includegraphics[totalheight=5cm,width=7cm]{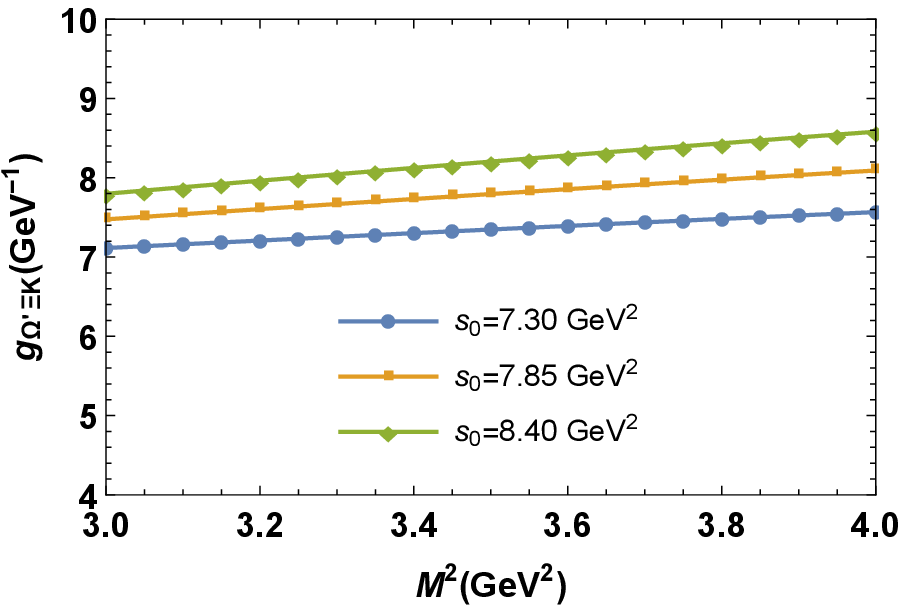}
\includegraphics[totalheight=5cm,width=7cm]{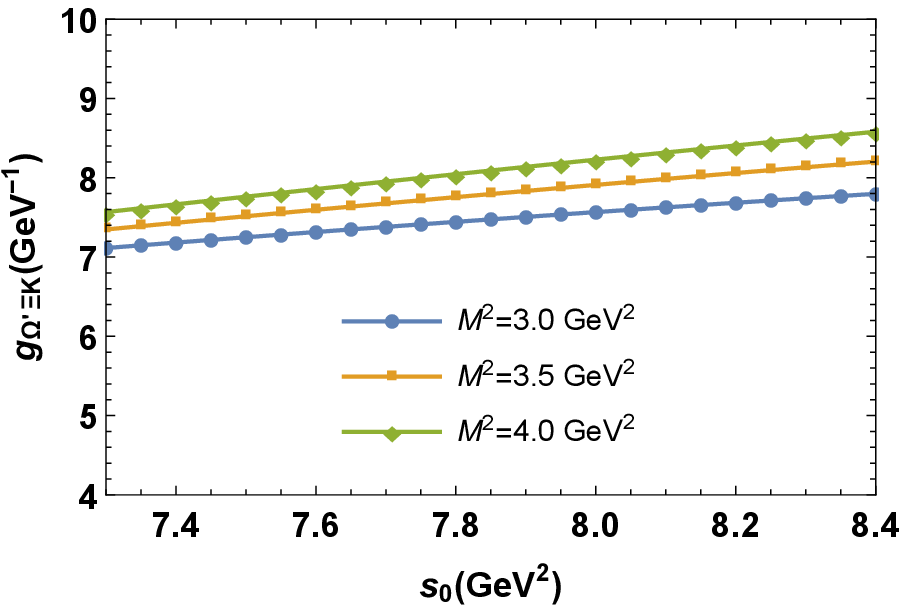}
\end{center}
\caption{\textbf{Left:} The coupling constant $g_{\Omega^{\prime}\Xi K}$ for the transition of the radially excited $\Omega$  baryon vs Borel parameter $M^2$ for the structure $ \slashed q q_{\mu} $.
\textbf{Right:} The coupling constant $g_{\Omega^{\prime}\Xi K}$ for the transition of the radially excited $\Omega$   baryon vs threshold
parameter $s_0$ for the structure $ \slashed q q_{\mu} $. }
\label{gr:gOmega2PXiK}
\end{figure}

The sum rules for the coupling constants for considered decays contain three auxiliary parameters, namely Borel mass $M^2$, continuum threshold $s_0$ and arbitrary parameter $\beta$ appearing in the expression of the interpolating current of $\Xi$ baryon. Since these are helping parameters, we need to find regions of these parameters, where physical quantities are practically independent of these parameters in their working regions. The lower limit of the $M^2$ is determined by demanding the convergence of the operator product expansion. The upper limit of the $M^2$ is determined by requiring pole dominance over the high states and continuum. Using these requirements the working region of the Borel parameter $M^2$ is adjusted as $M^2 \in [3,4]~\mbox{GeV}^2$. The continuum threshold $s_0$ is not arbitrary and it is related to the energy of the first excited state in initial channel. For determination of the working region of $s_0$ we demand that the result for the coupling constants change 10\%. It leads to $s_0 \in [7.3,8.4]~\mbox{GeV}^2$ region for $s_0$. Finally to determine the working region of the arbitrary parameter $\beta$ the dependence of the coupling constant on $\cos\theta$, where $\beta=\tan\theta$, is considered and looking upon the region with mild dependency on that parameter corresponding intervals are derived as $\cos\theta \in [-0.9,-0.5]$ and $\cos\theta \in [0.5,0.9]$. Using these intervals of auxiliary parameters, as examples, in Figs.~\ref{gr:gOmega1PXiK},  and \ref{gr:gOmega2PXiK} we depict the variations of the coupling constants  $g_{\widetilde{\Omega}\Xi K} $ and $g_{\Omega^{\prime}\Xi K}$ for the structure $ \slashed q q_{\mu} $ as a function of Borel mass $M^2$ and threshold parameter $s_0$. As is seen from the figures, though being not completely independent, the coupling constants show moderate dependency on the auxiliary parameters $M^2$ and $s_0$ which is acceptable in the error limits of the QCD sum rule formalism.  

With the parameters given in table~\ref{tab:Param1} and the determined working intervals of the auxiliary parameters, the values of the coupling constants for the interested transitions are obtained as presented in table~\ref{tab:DecayWidths}.
\begin{table}[tbp]
\begin{tabular}{|c|c|c|c|}
\hline\hline
 Structure& Decay & g~$(\mbox{GeV}^{-1})$ & $\Gamma$~(MeV) \\ \hline\hline
$\slashed q q_{\mu}$ & $\widetilde{\Omega}\rightarrow \Xi K$& $11.13^{+1.14}_{-1.31}$  & $6.98\pm 1.75$ \\
\hline
$\slashed q q_{\mu}$ &$\Omega^{\prime} \rightarrow \Xi K$& $7.76^{+0.71}_{-0.87}$   & $355.02\pm 88.68$ \\
\hline
$\slashed p q_{\mu}$ & $\widetilde{\Omega}\rightarrow \Xi K$& $11.57^{+1.42}_{-1.53}$  & $7.57
\pm 1.98$ \\
\hline
$\slashed p q_{\mu}$ &$\Omega^{\prime} \rightarrow \Xi K$& $7.53^{+0.69}_{-0.85}$   &
$333.64\pm 83.79$ \\
\hline
$\slashed q \slashed p q_{\mu}$ & $\widetilde{\Omega}\rightarrow \Xi K$&
$10.64^{+0.98}_{-1.14}$  & $6.41\pm 1.57$ \\
\hline
$\slashed q \slashed p q_{\mu}$ &$\Omega^{\prime} \rightarrow \Xi K$& $7.56^{+0.68}_{-0.76}$
   & $336.36\pm 86.79$ \\
\hline $ q_{\mu}$ & $\widetilde{\Omega}\rightarrow \Xi K$&
$11.54^{+1.39}_{-1.47}$  & $7.53\pm 1.87$ \\
\hline $ q_{\mu}$ &$\Omega^{\prime} \rightarrow \Xi K$&
$7.95^{+0.81}_{-0.92}$
   & $371.99\pm 92.85$ \\
 \hline\hline
\end{tabular}%
\caption{Sum rules predictions for the strong coupling constants and widths corresponding to  the decays of the orbitally and radially excited $\Omega$ states to $\Xi$ and $K$ using different Lorentz structures.}
\label{tab:DecayWidths}
\end{table}
The presented errors  arise from the uncertainties coming from the ones exist in the input parameters, as well as the uncertainties coming from the determination of the auxiliary parameters. 
 From table~\ref{tab:DecayWidths} we see that all Lorentz structures give close results for the coupling constants and they are consistent within the errors.  
With these coupling results we also calculate the decay widths corresponding to both the possible scenarios for excitation of ground state $\Omega$. Using the matrix elements for $\widetilde{\Omega}\rightarrow \Xi K$ and $\Omega^{\prime}\rightarrow \Xi K$ transitions, for decay widths we have
\begin{eqnarray}
\Gamma (\widetilde{\Omega} &\rightarrow &\Xi K)=\frac{%
g_{\widetilde{\Omega}\Xi K}^{2}}{24\pi m_{\widetilde{\Omega}}^{2}}\left[(m_{\widetilde{\Omega}}%
-m_{\Xi })^{2}-m_{K}^{2}\right]  f^3(m_{\widetilde{\Omega}},m_{\Xi},m_{K}),
\end{eqnarray}%
\begin{eqnarray}
\Gamma (\Omega ^{ \prime } &\rightarrow &\Xi K)=\frac{%
g_{\Omega ^{\prime }\Xi K}^{2}}{24\pi m_{\Omega^{\prime}}^2}\left[ (m_{\Omega^{\prime}
}+m_{\Xi})^{2}-m_{K}^{2}\right]  f^3(m_{\Omega^{\prime}},m_{\Xi},m_{K}),
\end{eqnarray}
respectively. The function $f(m_{\widetilde{\Omega}(\Omega^{\prime})},m_{\Xi},m_{K})$ has the form
\begin{eqnarray}
f(m_{\widetilde{\Omega}(\Omega^{\prime})},m_{\Xi},m_{K})=\frac{1}{2m_{\widetilde{\Omega}(\Omega^{\prime})}}\sqrt{m_{\widetilde{\Omega}(\Omega^{\prime})}^4+m_{\Xi}^4+m_K^4-2m_{\widetilde{\Omega}(\Omega^{\prime})}^2m_{\Xi}^2-2m_{\widetilde{\Omega}(\Omega^{\prime})}^2m_K^2-2m_{\Xi}^2m_K^2}.
\end{eqnarray}
Using the values of the coupling constants obtained for each possible case, the results for the decay widths are obtained as also presented in table~\ref{tab:DecayWidths} for different Lorentz structures. As is again seen, the numerical values of the decay widths for $\widetilde{\Omega}\rightarrow \Xi K$ and $\Omega^{\prime} \rightarrow \Xi K$ obtained using different structures are consistent  within the errors.

Before comparison of the results with the experimental data we need to remark that in the present work we calculate the partial width of the  $\widetilde{\Omega}\rightarrow \Xi K$ and $\Omega^{\prime} \rightarrow \Xi K$ transitions. The Belle experiment has observed $\Omega(2012)$ resonance in the $\Omega^{*-}\rightarrow \Xi^0K^-$ and $\Omega^{*-}\rightarrow \Xi^{-}\bar{K}^0$ modes and measured the ratio of the branching fractions in these channels, $ {\cal R}=\frac{{\cal B}(\Omega^{*-}\rightarrow \Xi^0K^-)}{{\cal B}(\Omega^{*-}\rightarrow \Xi^-\bar{K}^0)} =1.2 \pm 0.3$ \cite{Yelton:2018mag}. There are no other experimentally observed two-body decays of this particle. Its possible three-body decay modes are expected to have small contributions in the framework of QCD sum rules. The main idea of QCD sum rule method is the dominance of the single-particle states over the states with two or more particles in phenomenological parts of sum rules. For calculation of three-particle (but not semileptonic) decay modes we face with two-particle states' contribution in phenomenological part of QCD sum rules. Estimation of such type contribution is very problematic in this method  and even if it is possible it should be suppressed via phase volume and the sum rules are not reliable. Hence, the two modes $\Omega^{*-}\rightarrow \Xi^0K^-$ and $\Omega^{*-}\rightarrow \Xi^{-}\bar{K}^0$ can be considered as dominant modes of the particle under consideration. Considering this point and the above ratio, we provide the total widths of the states under consideration for all the structures in table \ref{tab:DecayWidthstot}.
\begin{table}[tbp]
\begin{tabular}{|c|c|c|}
\hline\hline
 Structure& State &  $\Gamma_{tot}$~(MeV) \\ \hline\hline
$\slashed q q_{\mu}$ & $\widetilde{\Omega}$  & $12.80\pm 3.21$ \\
\hline
$\slashed q q_{\mu}$ &$\Omega^{\prime} $  & $650.75\pm 162.55$ \\
\hline
$\slashed p q_{\mu}$ & $\widetilde{\Omega}$  & $13.87
\pm 3.62$ \\
\hline
$\slashed p q_{\mu}$ &$\Omega^{\prime} $  &
$611.56\pm 153.58$ \\
\hline
$\slashed q \slashed p q_{\mu}$ & $\widetilde{\Omega}$  & $11.74\pm 2.87$ \\
\hline
$\slashed q \slashed p q_{\mu}$ &$\Omega^{\prime}$
   & $616.54\pm 159.08$ \\
\hline $ q_{\mu}$ & $\widetilde{\Omega}$  & $13.80\pm 3.42$ \\
\hline $ q_{\mu}$ &$\Omega^{\prime} $
   & $681.85\pm 170.19$ \\
 \hline\hline
\end{tabular}%
\caption{Total widths corresponding to   the orbitally and radially excited $\Omega$ states.}
\label{tab:DecayWidthstot}
\end{table}
Comparing the results obtained for the total widths of the considered states  with the experimental value of width,  $6.4^{+2.5}_{-2.0}$~(stat)~$\pm 1.6$~(syst)~MeV~\cite{Yelton:2018mag}, we see that the results for orbitally excited states are consistent with the experimental width within the errors for all structures. The total widths obtained for the radially excited states, however, are very far from the experimental value of the width for all structures. With these information, we conclude that the newly observed $\Omega(2012)$ is $1P$ excitation of the ground state $\Omega$ and this assignment is structure independent. Among all Lorentz structures the $\slashed q \slashed p q_{\mu}$, containing maximum number of momenta, gives closer  value to that of the experimental width compared to other structures.

\section{Conclusion}
Using the LCSR method we made a calculation on the coupling constant and decay width of the $\Omega(2012)$ transition to $\Xi K$ considering it as  either orbital  or radial excitation in $ \Omega $ channel.  In a previous work, \cite{Aliev:2018syi}, we extracted the mass of the $\Omega(2012)$ taking it as an orbital excitation of the ground state $ \Omega $  baryon and got a result in a good consistency with the experimentally measured value. In Ref.~\cite{Aliev:2016jnp} the corresponding mass was calculated for the radial excitation of the state under consideration  and obtained as $2176\pm 219$~MeV. Comparison of these results with the experimentally obtained mass allowed us to reach a conclusion on the quantum numbers of the observed $\Omega(2012)$ state as $J^P=\frac{3}{2}^-$. The present work has been done to gain more new information on this issue. Hence, the decay widths were calculated for the channels $\widetilde{\Omega}\rightarrow \Xi^0K^-$ and $\Omega^{\prime}\rightarrow \Xi^0K^-$  with the assumption of $\Omega(2012)$ being an orbital or radial excitation of the ground state $\Omega$. Considering the $\Omega^{*-}\rightarrow \Xi^0K^-$ and $\Omega^{*-}\rightarrow \Xi^{-}\bar{K}^0$ as dominant modes  of $\Omega(2012)$ state and  the measured ratio $ {\cal R}=\frac{{\cal B}(\Omega^{*-}\rightarrow \Xi^0K^-)}{{\cal B}(\Omega^{*-}\rightarrow \Xi^-\bar{K}^0)} =1.2 \pm 0.3$ \cite{Yelton:2018mag} we estimated the total widths of the considered states for all the Lorentz structures. Comparing the obtained results for the total widths with the observed width value reported by the Belle Collaboration, $6.4^{+2.5}_{-2.0}$~(stat)~$\pm 1.6$~(syst)~MeV~\cite{Yelton:2018mag}, we concluded that $\Omega(2012)$ is $1P$ excitation of the ground state $\Omega$. This conclusion, which is structure independent, supports our previous assignments for the nature of $\Omega(2012)$ resonance using mass sum rules. Hence,  the $\Omega(2012)$ state is first orbital excitation in $ \Omega $ channel with spin-parity $J^P=\frac{3}{2}^-$. This assignment for the spin-parity of $\Omega(2012)$ is consistent with those of Refs.~\cite{Polyakov:2018mow,Valderrama:2018bmv,Lin:2018nqd}, which consider this particle as the $ \Xi^* K$, $ \Xi(1530) K$ or $ \Xi(1530) \bar{K} $ molecular state.  In the molecular picture considered in  Refs.~\cite{Polyakov:2018mow,Valderrama:2018bmv,Lin:2018nqd} the three-body decays take place at tree level, but the two-particle decays appear only at loop level. Therefore,  in this picture, the three-body decays dominate over the two-body modes. In our case that we consider $\Omega(2012)$ as a usual three-quark state, however, as we also noted before any estimation on the three-body decays of $\Omega(2012)$ is problematic regarding the general philosophy of the QCD sum rule method and  even if it becomes possible it should be suppressed via phase volume. Therefore, more measurements on the possible  two- and three-body decay modes of $\Omega(2012)$ state  can play decisive roles on choosing the  ``right" picture for the internal structure of  this particle.

\section*{ACKNOWLEDGEMENTS}

H. S. thanks Kocaeli University for the partial financial support through the grant BAP 2018/070.

\label{sec:Num}

\end{document}